\documentclass[preprint]{sigchi}

\usepackage{balance}       % to better equalize the last page

\usepackage{wrapfig}
\usepackage{graphics}      % for EPS, load graphicx instead 
\usepackage[T1]{fontenc}   % for umlauts and other diaeresis
\usepackage{sidenotes}
\usepackage{mathptmx}
\usepackage[pdflang={en-US},pdftex]{hyperref}
\usepackage{color}
\usepackage{booktabs}
\usepackage{textcomp}

\usepackage{microtype}        % Improved Tracking and Kerning
\usepackage{ccicons}          % Cite your images correctly!

\usepackage{todonotes}
\newcommand{\com}[1]{}

\newcommand{\old}[1]{}

\def\plaintitle{StripBrush: A Constraint-Relaxed 3D Brush Reduces Physical Effort and Enhances the Quality of Spatial Drawing}

\def\plainauthor{Enrique Rosales, Jafet Rodriguez, Chrystiano Ara\'ujo, Nicholas Vining, Dongwook Yoon, Alla Sheffer}

\def\plainkeywords{Spatial drawing; VR drawing; 3D sketching; constraint relaxation; 3D brush design; wrist-twisting motion.}

\makeatletter
\def\url@leostyle{%
  \@ifundefined{selectfont}{
    \def\UrlFont{\sf}
  }{
    \def\UrlFont{\small\bf\ttfamily}
  }}
\makeatother
\def\pprw{8.5in}
\def\pprh{11in}

\definecolor{linkColor}{RGB}{6,125,233}
\hypersetup{%
  pdftitle={\plaintitle},
  pdfauthor={\plainauthor},
  pdfsubject={Computer Science, Human-Computer Interaction},
  pdfkeywords={\plainkeywords},
  pdfdisplaydoctitle=true, % For Accessibility
  bookmarksnumbered,
  pdfstartview={FitH},
  colorlinks,
  citecolor=black,
  filecolor=black,
  linkcolor=black,
  urlcolor=linkColor,
  breaklinks=true,
  hypertexnames=false
}

\begin{document}

\title{\plaintitle}

\numberofauthors{6}
\author{%
  \alignauthor{Enrique Rosales\\
    \affaddr{University of British Columbia, Canada}\\
    \affaddr{Universidad Panamericana, M\'{e}xico}\\
    \email{albertr@cs.ubc.ca}}\\
  \alignauthor{Jafet Rodriguez\\
    \affaddr{Universidad Panamericana, M\'{e}xico}\\
    \email{arodrig@up.edu.mx}}\\
  \alignauthor{Chrystiano Ara\'ujo\\
    \affaddr{University of British Columbia, Canada}\\
    \email{araujoc@cs.ubc.ca}}\\   
  \alignauthor{Nicholas Vining\\
    \affaddr{University of British Columbia, Canada}\\
    \affaddr{NVIDIA, Canada}\\
    \email{nvining@cs.ubc.ca}}\\
  \alignauthor{Dongwook Yoon\\
    \affaddr{University of British Columbia, Canada}\\
    \email{yoon@cs.ubc.ca}}\\
  \alignauthor{Alla Sheffer\\
    \affaddr{University of British Columbia, Canada}\\
    \email{sheffa@cs.ubc.ca}}\\
}

%\toappear{2021 Copyright held by the authors.}
\maketitle

\begin{abstract}
Spatial drawing using ruled-surface brush strokes is a popular mode of content creation in immersive VR, yet little is known about the usability of existing spatial drawing interfaces or potential improvements. We address these questions in a three-phase study. (1) Our exploratory need-finding study (N=8) indicates that popular spatial brushes require users to perform large wrist motions, causing physical strain. We speculate that this is partly due to constraining users to align their 3D controllers with their intended stroke normal orientation. (2) We designed and implemented a new brush interface that significantly reduces the physical effort and wrist motion involved in VR drawing, with the additional benefit of increasing drawing accuracy. We achieve this by relaxing the normal alignment constraints, allowing users to control stroke rulings, and estimating normals from them instead. (3) Our comparative evaluation of StripBrush (N=17) against the traditional brush shows that StripBrush requires significantly less physical effort and allows users to more accurately depict their intended shapes while offering competitive ease-of-use and speed.
\end{abstract}

% Author Keywords

\section{Introduction}
\label{sec:intro}

Spatial drawing is an increasingly popular mode of content creation in immersive VR. Virtual reality device manufacturers have provided art packages for their hardware \cite{TiltBrush:2019,OculusMedium:2019}  targeted at both amateurs 
and professionals, as a competitive selling point.  
Despite the popularity and potential of spatial sketching applications, and the recent resurgence of interest in virtual reality software and hardware at a consumer level, numerous barriers prevent true widespread commercial adoption. 
Three-dimensional interfaces, and virtual reality devices in general, cause new usability challenges without precedent in traditional screen-based GUIs, including "gorilla arm syndrome", and depth perception problems. However, HCI and VR literature lags behind software 
development in understanding the usability challenges of VR haptic interfaces specific to 3D drawing tasks, and ways of addressing them.

\setlength{\abovecaptionskip}{3pt}
\begin{figure}
\centering
  \includegraphics[width=\linewidth]{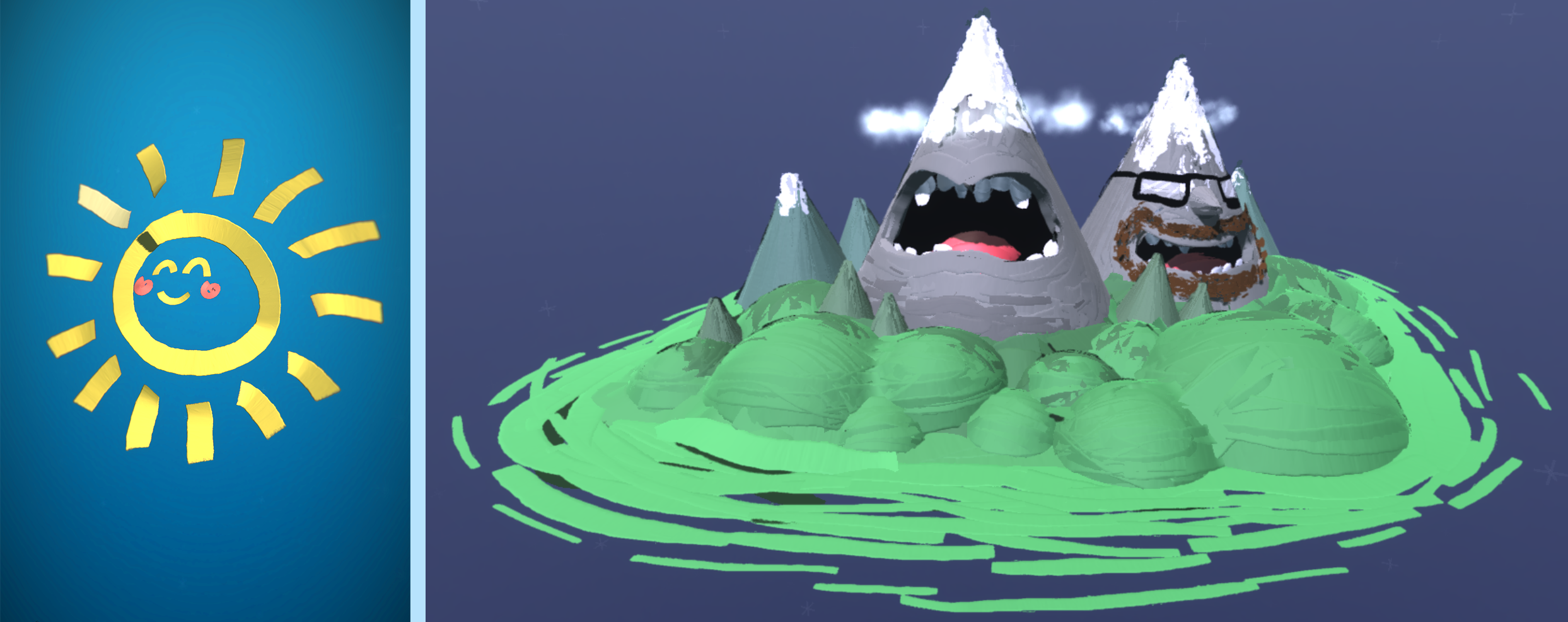}
\caption{Representative ribbon brush drawings. Left: $\copyright$ Enrique Rosales, Right: $\copyright$ Andrew Bell.}
\label{fig:drawings}
\vspace{-5pt}
\end{figure}

\setlength{\abovecaptionskip}{3pt}
\begin{figure}
\centering
  \includegraphics[width=\linewidth]{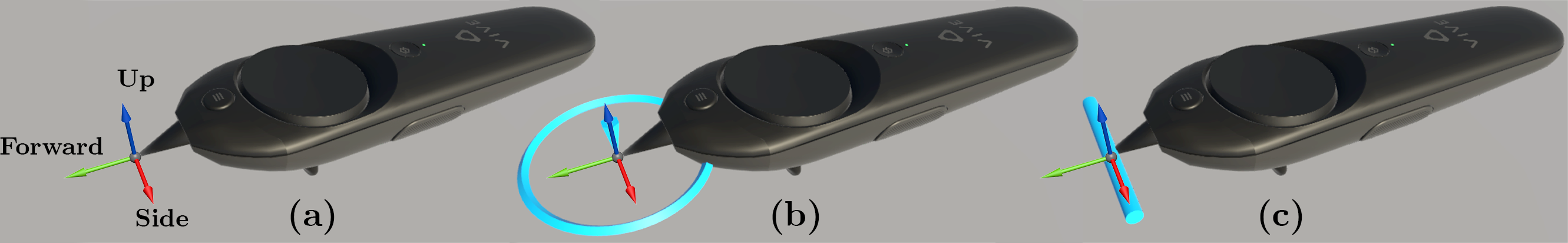}
\caption{ (a) Canonical VR controller coordinate system. (b,c) Ribbon brush interfaces: the normal brush interface uses the `up' axis as proxy for ribbon normal, and expects users to align a circular disk with their intended ribbon tangent plane; (c) our StripBrush interface directly employs the `side' axis as the ribbon ruling. }
\label{fig:brush}
\label{fig:controller}
\vspace{-10pt}
\end{figure}

We study these drawing-specific usability challenges, focusing on popular ruled-surface\footnote{A {\em ruled surface} is a surface that can be swept out by moving a line, or {\em ruling}, in space.}, or {\em ribbon} based, brushes (Figure~\ref{fig:drawings}), and consider the mechanisms behind how users communicate their intended ribbon shapes - specifically, the ribbon's spatial orientation or normal. 

One of the most popular implementations of 3D brushes - hereafter referred to as the {\em normal brush} - uses the orientation of the user's controller at drawing time to define the orientation of the ribbon surface, as illustrated in Figure~\ref{fig:brush}(b) (specifically, it uses the `up' direction of the controller's local coordinate frame as a proxy for the user-intended ribbon normal). Our aim is to critically examine the design of this interface, with likely the largest outreach in VR drawing software. 
We therefore commence this study by asking the following research questions:

\setlength{\abovecaptionskip}{3pt}
\begin{figure}
\centering
  \includegraphics[width=\linewidth]{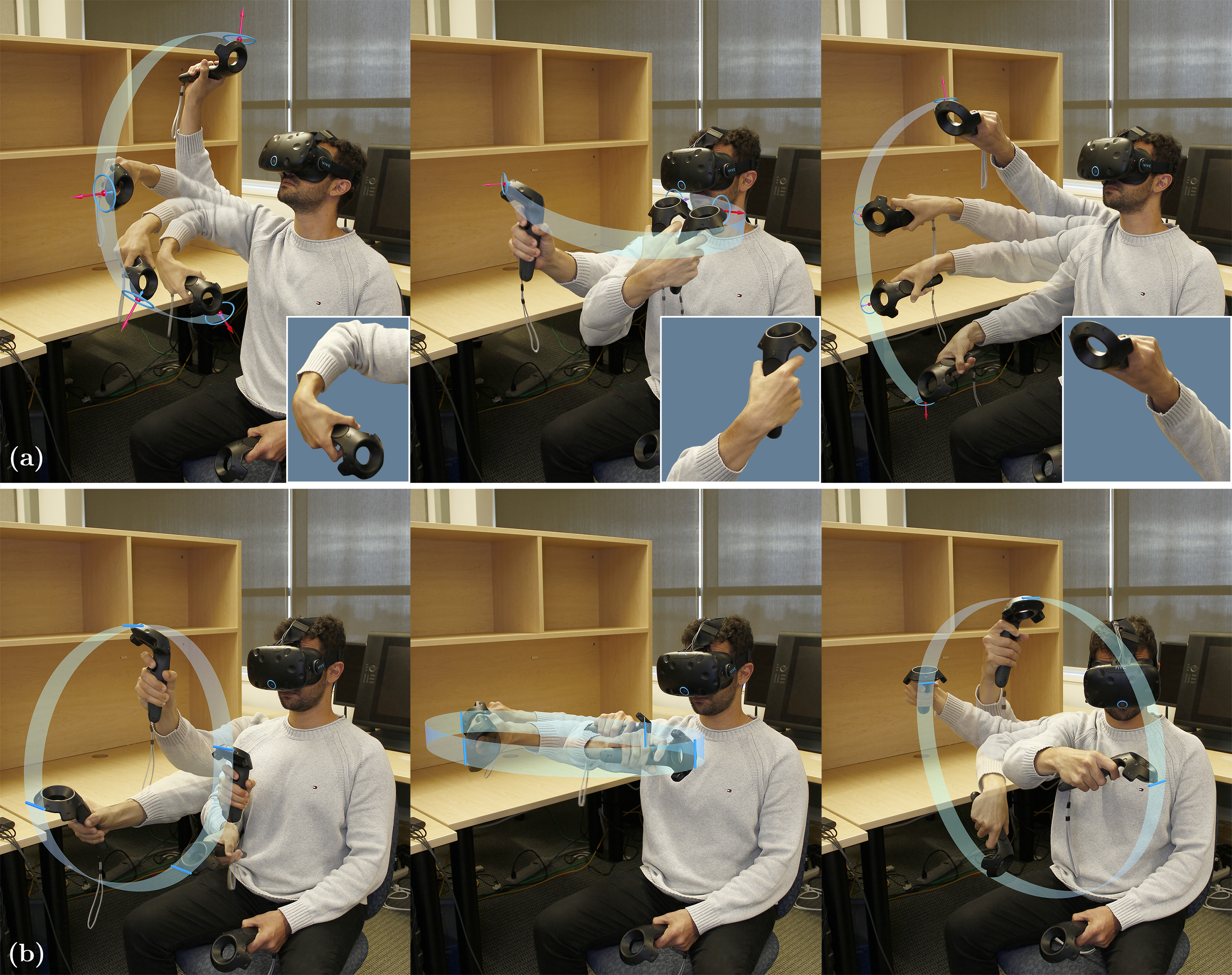}
\caption{Using the normal brush (a) can require users to twist their hands into awkward positions; using StripBrush can (b) significantly reduce users' physical discomfort.}
\label{fig:painful}
\vspace{-10pt}
\end{figure}

\begin{enumerate}
\item RQ1. What are the user's needs and frustrations, if any, specific to the design of the normal brush when the users draw different shapes in VR?

\item RQ2. How do we design and implement a new brush interface that can address the identified normal brush problems?

\item RQ3. To what extent can our design solution address the problems posed by the normal brush?
\end{enumerate}

To answer these questions, we performed an explorative needs-finding study (N=8) in which we asked users to surface predetermined shapes that span a wide range of surface curvature patterns. We then analyzed their recorded motions and their feedback. In this study, we identified wrist-twisting motions as an important usability problem specific to spatial ribbon-stroke drawing tasks. The normal brush used by virtual reality ribbon drawing programs constrains users to align their 3D controller, and hence the hand that holds it, with their intended stroke normal direction; we found that this often forced users to twist their hands into awkward positions when drawing curved surfaces best depicted using strokes with high normal variation (Figure~\ref{fig:painful}(a)). 
We characterized the source of this wrist-twisting problem as due to the constrained nature of the controller-to-surface mapping used by the normal brush.

We suggest constraint relaxation as a novel and effective solution to the over-constraining problem of the normal brush. After analyzing the mathematical properties of ruled surfaces and the possible linkages 
between controller orientation and resulting ruled surface shapes, we first replaced the over-constrained ruled surface definition employed by the normal interfaces with a more general one (specifically, we relaxed the implicit requirement, used by normal brushes, for rulings to be orthogonal to their sweeping trajectory). We then explored the core design spaces for ideating and designing new ways to control ribbon ruling orientations. We converged to a control metaphor, the {\em StripBrush}, that has users directly specify rulings using a natural mapping from everyday painting tools (Figure~\ref{fig:brush}(c)) and significantly reduces physical discomfort (Figure~\ref{fig:painful}(b)). 

Our comparative evaluation of StripBrush (N=17) against the normal brush shows that StripBrush requires significantly less physical effort and allows users to more accurately depict their intended shapes while offering competitive ease-of-use and speed.

This paper makes three major contributions: (1) identification and characterization of wrist-twisting motions as a core usability challenge of ribbon-based 3D drawing interfaces, (2) the design of StripBrush and its technical implementation, and (3) evaluation results which are indicative of the efficacy of StripBrush and have design implications for further improvement.

\section{Related Work}
\label{sec:relatedWork}
\label{sec:prev}
Our work is inspired by prior research on  interfaces for 3D modeling, and 2D or 3D curve drawing, and by ergonomics of VR-based interfaces in general and drawing interfaces in particular.

\subsection{Shape Modeling Interfaces}

Research on interfaces for modeling or creating 3D content can be traced back to the early SketchPad system of Sutherland~\shortcite{Sutherland:1964}. The range of state-of-the-art modeling interfaces spans from traditional menu-based systems such as Maya 
\shortcite{Maya} or 3D Studio Max ~\cite{3DSMax}, through sculpting based systems such as ZBrush \cite{ZBrush}, to 2D sketch-based interfaces ~\cite{Olsen:2009,Jorge:2011} and VR-based tools such as 
TiltBrush~\cite{TiltBrush:2019}, GravitySketch~\cite{GravitySketch:2019}, Oculus Medium~\cite{OculusMedium:2019}, and Google Blocks~\cite{GoogleBlocks:2019}. While menu-based and sculpting-based systems primarily target expert modelers, many 2D-sketch-based modeling 
frameworks aim for a broader user base. However despite recent advances~\cite{Nealen:2007,Xu:2014,Igarashi:1999,Bae:2008,Li:2017}, their scope is still limited. 

There are various genres of VR-based modeling tools. VR sculpting tools \cite{Kodon:2019,OculusMedium:2019,ShapeLab:2019} target expert users interested in creating complex free-form shapes and aim for a more fluid interaction than their traditional counterparts. 
VR systems that target target low-expertise users often allow users to create simple CSG shapes by applying Boolean operations to a fixed set of initial primitives \cite{DesignSpace:2019,Diehl:2004,GoogleBlocks:2019,Tano:2013}. 
Researchers have proposed several VR interfaces for drawing swept surfaces in 3D space~\cite{Keefe:2001,Schkolne:2001}, an approach that has been popularized by the GravitySketch commercial system \cite{GravitySketch:2019}. Using such interfaces to 
effectively create more general geometry requires non-trivial expertise \cite{Rosales:2019}. 

Another option, popularized by several VR interfaces, is to use 3D sketches as a stepping stone toward generating free-form 3D shapes. Earlier VR sketch-based modeling systems target users with some design or modeling expertise, and expect these users to draw
connected 3D curve networks~\cite{Fiorentino:2002,Jackson:2016,Kwon:2005,Wesche:2001,Sachs:1991}; these networks can then be surfaced algorithmically by employing generic~\cite{Farin:2001} or more targeted~\cite{Bessmeltsev:2012,Pan:2015} surfacing methods. 
State-of-the-art methods~\cite{Rosales:2019,Huang:2019} directly convert dense ribbon-brush VR drawings consisting of disjoint stroke collections into manifold surfaces. In this study, we seek to lower the barriers to VR-based freeform sketching 
tools, with a keen focus on the design of spatial brushes at the forefront of such tools intended for a wide variety of users, including both novices and experienced users.

\subsection{Pen and Brush Interfaces}

2D stylus-based stroke drawing interfaces are increasingly ubiquitous, migrating from high-end hardware~\cite{Wacom} to consumer-level laptops and tablets. Raw user strokes are captured as polylines replicating the path of the tip of the user's stylus over the drawing pad 
or a pen-sensitive display. Software systems that use this raw input, such as Adobe Illustrator~\cite{AdobeIllustrator}, typically apply local fairing~\cite{Baran:2010,McCrae:2009} to the recorded path data to overcome user hand-tremors and other sources of noise in the 
recorded path. 

Researchers have experimented with a range of techniques to form three-dimensional curves~\cite{Israel:2009,Grossman:2002,Tano:2003,Jackson:2016,Diehl:2004,Amores:2017,Kim:2018} to enable direct 3D drawing. 
For example, Grossman et al. \cite{Grossman:2002} allow users to manipulate a 
physical tape that represents the 3D curve; Keefe et al. \cite{Keefe:2007} propose a system consisting of a virtual pencil, attached to a haptic device, and a stereoscopic display. Multiple VR systems \cite{TiltBrush:2019,PaintLab:2019,Quill,GravitySketch:2019,Keefe:2001,Keefe:2007} allow users to draw 3D strokes that follow the path of a hand-held controller  and are rendered in real time, providing users with instant visual feedback. 
These tools are becoming increasingly popular, as they allow users to create rich  3D content (Figure~\ref{fig:drawings}) and are well suited for users to express and communicate purely virtual, static visuals of their intended shapes. 
Many systems~\cite{PaintLab:2019,Keefe:2007} represent and render the strokes as 
tubular shapes centered around the captured tip paths. Others~\cite{Keefe:2001,Keefe:2007,TiltBrush:2019,Quill,GravitySketch:2019} generate ribbon-like  strokes, defined by sweeping a ruling line along the captured controller path.

Our study focuses on these ribbon-based approaches, for the promise they show in not only creating rich 3D visuals but also serving as intermediate data for  3D  surface modeling~\cite{Rosales:2019}. Uniquely defining the shape of a ruled surface ribbon requires specifying the 
orientation of the ribbon's rulings at each point along its trajectory. Existing interfaces \cite{Keefe:2001,Keefe:2007,TiltBrush:2019,Quill} interpret the orientation of the hand-held controller as a proxy for the surface normal, as described in 
Section~\ref{sec:explorativeStudy} and use this normal to derive a ruling orientation. These systems convey this linkage to the users by rendering a circular disk at the tip of the controller, aligning the disk normal with the ''up'' orientation of the controller's coordinate system. (Figure~\ref{fig:brush}, (a,b)). As our study 
(Section~\ref{sec:explorativeStudy}) shows, using this circle-based interface requires users to exert significant flexion and torque through the wrist and elbow joints to create their desired drawings; additionally, as discussed in Section~\ref{sec:explorativeStudy} the planar circle may provide misleading feedback to the 
user as to the geometry of the output ribbon, as the actual normal may significantly diverge from the user-entered proxy. Very recently (September 2019) GravitySketch unofficially pre-released a ribbon-drawing tool that has users directly specify the ribbon rulings by 
positioning and moving a line segment rendered along the ''forward'' axis  of the controller's coordinate system. We compare our brush design against this alternative in Section~\ref{sec:design} and explain design rationales for our approach.

\subsection{Ergonomics of generic 3D VR Interfaces}

Previous work in HCI and VR literature has thoroughly documented the problem of "gorilla arm" \cite{laviola3D2017}, in which "mid-air interactions are prone to fatigue and lead to 
a feeling of heaviness in the upper limbs" \cite{Ramos2014}. Fatigue modeling \cite{Ramos2014, Jang2017} provides tools for estimating subjective fatigue caused by gorilla arm, but primarily focuses on estimating mid-air pointing fatigue as a function of shoulder joint 
torque, and does not evaluate strain caused by torque applied to the wrist. Anecdotal evidence exists that overconstrained VR interfaces can cause physical strain \cite{Grossman2003, Zhai1998}. We conceptualize the wrist-twisting behavior for brush-based VR interfaces as 
caused by requiring users to provide {\em over-constrained} input that requires unnatural exertion (Section~\ref{sec:explorativeStudy}). 
This observation serves as the basis for suggesting constraint relaxation as a solution (Section~\ref{sec:design}).

\subsection{Empirical Studies of VR Drawing} 
A recent study \cite{Arora:2017} noted that four of five expert users experienced ergonomic issues such as neck and shoulder pain when using a VR sketching system over 
an extended period. Several studies analyzed user accuracy when drawing in space, concluding that 3D drawing is far less accurate than its 3D counterpart~\cite{Keefe:2007,Rauch:2010,Arora:2017}. Arora et al. \cite{Arora:2017} indicate that 3D drawings are rarely accurate due 
to lack of a physical drawing surface, and that visual guidelines are insufficient to improve accuracy. Our comparative study (Section~\ref{sec:evaluation}) shows how different 3D brush designs impact user-perceived drawing accuracy and their physical strain. Barera Machuca 
et al. \cite{Machuca2019} suggest that user's spatial ability influences shape quality. Wiese et al.~\cite{Wiese:2010} demonstrate that users VR drawing skills improve rapidly as they gain more experience.  Our study (Section~\ref{sec:evaluation}) reinforces this conclusion 
for the scenarios tested.

\begin{figure}
\centering
  \includegraphics[width=.8\linewidth]{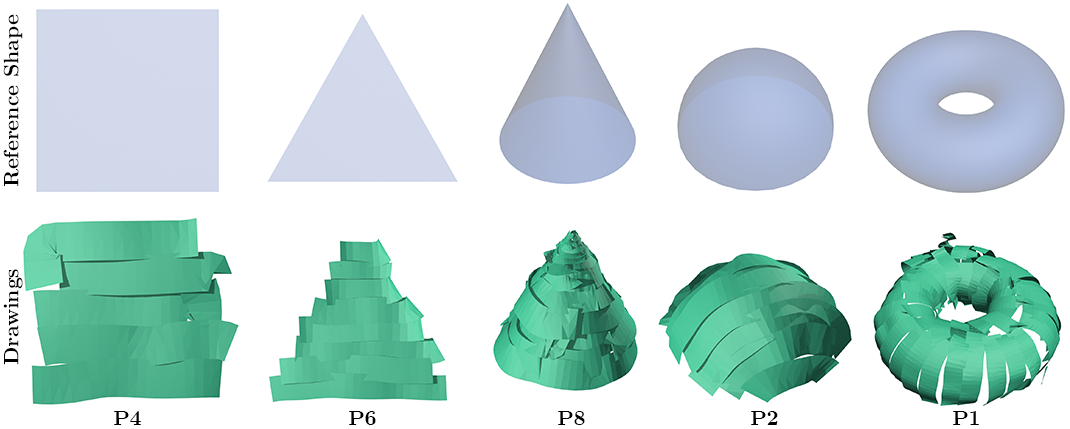}
\caption{Formative study: (top) test shape set. (bottom) representative participant 3D drawings of these shapes.}
\label{fig:expl_results}
\end{figure}

\section{Formative Evaluation of Traditional Spatial Brush Interfaces}
\label{sec:explorativeStudy}
We performed a formative study to examine how users worked with the most widely-adopted implementation of spatial sketching interfaces, the \emph{normal-based} ribbon brush, and to identify areas where users encountered challenges and might benefit from interface 
improvements. Specifically, we observed how non-expert users drew a set of target shapes in VR using normal brushes when instructed to surface a target shape as cleanly as possible, and then sought feedback from the users about their experience. Our goal was to answer 
the following questions:

\begin{enumerate}
\item How challenging do users find the task of drawing using normal brushes?
\item What are the common challenges that stem from the design of normal-based VR drawing interfaces?
\item How does the target shape that the user is attempting to draw affect these challenges?
\end{enumerate}

\subsection{Apparatus: The Normal Brush}

Normal-based ribbon drawing interfaces present the user with a disk orthogonal to the controller's `up' direction (Figure~\ref{fig:brush},a). At drawing time, they record this direction as $n_c$, and the position of the controller tip as $p_c$. 

\setlength\columnsep{5pt}
\begin{wrapfigure}[4]{l}[0pt]{.34\linewidth}
\vspace{-15pt}
  \begin{center}
   \includegraphics[width=\linewidth]{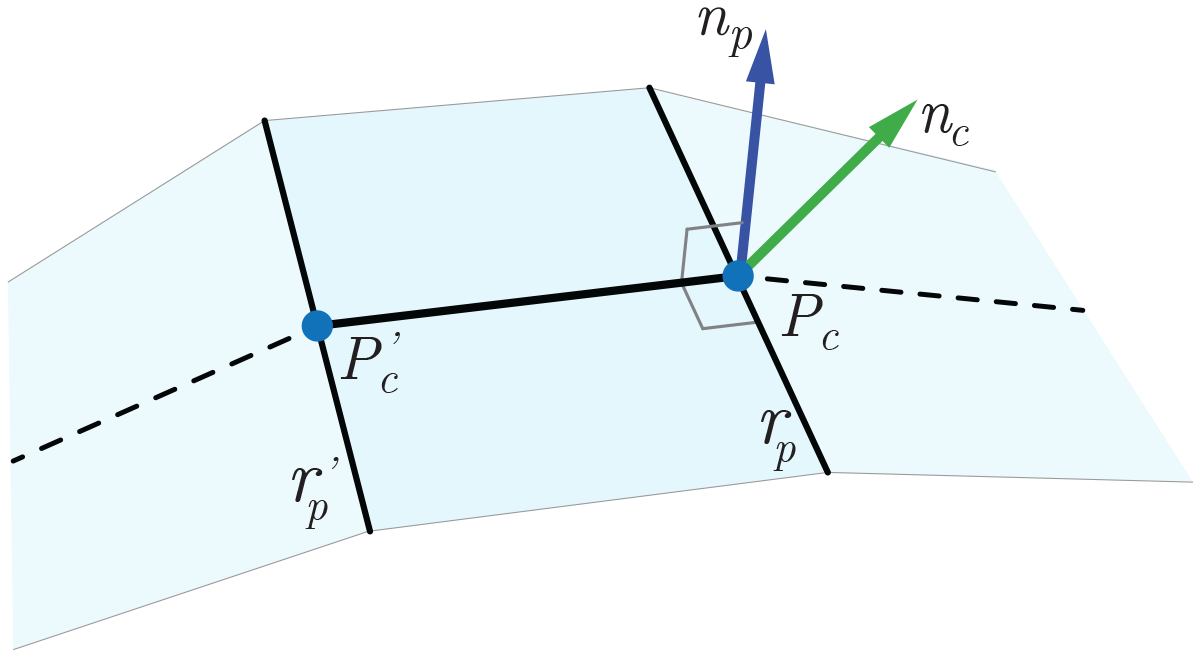}
  \end{center}
\end{wrapfigure}

To embed the controller's spatial path in the rendered ribbons, for each new position $c_p$ and given the previous captured position $p'_c$, they compute the direction of the ruling $r_p$ at $p_c$ as $r_p=n_c \times (p_c - p'_c)$ (see inset). They then form a discrete ruled surface comprised of 
quadrilaterals along the stroke path by connecting consecutive vertices along the left and right sides of the rulings. This construction generates rulings which are orthogonal to the vectors $\overrightarrow{p'_c p_c}$ but does not guarantee alignment between 
ribbon and controller normals (see difference between the ribbon normal $n_p$ at $p$ and $n_c$ in the inset); nor does it provide an upper bound on the angle between these input and output normals.

\subsection{Methodology}
The core promise of ribbon brush interfaces is their suitability for drawing free-form shapes, both closed and open, with a wide range of curvature configurations and magnitudes. Therefore, we tasked our participants with drawing a set of shapes (Figure~\ref{fig:expl_results}, top) that covers all possible curvature configurations (planar, spherical (isotropic non-planar), parabolic, elliptical, and hyperbolic)~\cite{doCarmo} and include both open and closed surfaces:

\begin{itemize}
\item{\textbf{Square}: Planar; $k_{min}= k_{max} = 0$, where $k_{min}$ and $k_{max}$ are minimal and maximal principal curvature values respectively).}
\item{\textbf{Triangle}: Planar;  $k_{min}= k_{max} = 0$.}
\item{\textbf{Cone}: Parabolic  $k_{min}=0, k_{max} > 0$.} 
\item{\textbf{Hemisphere}: Spherical $k_{min}= k_{max} > 0$.}
\item{\textbf{Torus}: Contains both elliptic ($k_{max}> k_{min} > 0$) and hyperbolic {$k_{min}> 0, k_{max} > 0$} regions at outer and inner parts of the torus}
\end{itemize}

Note that two shapes are planar (square and triangle); one is singly-curved, namely it has one non-zero, and one zero, principal curvatures (cone); and two are doubly curved, namely they have two non zero principal curvatures (hemisphere and torus). 
The torus is a closed shape, while the others are open.  

All participants were given a brief introduction to the drawing user interface (20 minutes) and a short explanation of the test (5 minutes). We then asked them to draw the five shapes. They were asked to draw each shape over a predefined scaffold, shown as a 
semi-transparent surface, replicating the underlying surfaces as cleanly as possible. Rosales et al.~\cite{Rosales:2019} indicate that first time users require time to achieve what they consider an acceptable drawing quality, but 
ultimately converge to a common drawing style, depicting surfaces using partially overlapping ribbon strokes with locally similar tangents. Thus to save user time, we suggested to the participants that they follow a similar style, placing strokes side-by-side and avoiding crisscrossing strokes. 
The order in which shapes were shown to the participants was balanced.

After drawing each shape, participants were asked to express their level of agreement with the statement that ``Drawing this surface is hard'' using a 1 to 5 scale (strongly disagree-strongly agree, Likert-style questions). Finally, we conducted follow-up semi-structured interviews (15 min) with a focus on specific usability challenges they encountered, if any.

The task (interview) sessions were video- (audio-) recorded for future analysis. We manually annotated our video recordings, marking every time the users erased a previous stroke, or performed an undo operation. We counted the number of these {\em error corrections}, per 
participant and per shape. A detailed record of the answers and the error correction counts ($8 \times 5 \times 2$) is provided in the supplementary material and is summarized in Figure \ref{fig:histograms} and Table \ref{tab:table1}.

\paragraph{Sampling and Screening} Prior research \cite{Arora:2017} indicates that a small percentage of the population lacks perception of depth in VR, and thus faces an inherent, tool-independent, difficulty when using VR content creation tools. Since our goal is to focus 
on challenges presented by a specific interface, rather than globally, we identified such participants and excluded them from the study using the following test: we asked the participants to draw three vertical ribbons side-by-side and roughly at the same depth. Two 
participants (one male and one female) were unable to place the ribbons at the same depth by far (depth variation was larger than three times the ribbon width) even after several attempts, and were excluded. Through convenience sampling, we recruited eight participants (6 male, 
2 female, all passed the depth-perception screening) consisting of 6 CS graduate students and 2 graduate students in electrical engineering. One had previous experience with VR drawing.

\subsection{Results and Discussion}

Overall, the participant-generated drawings captured the core geometric characteristics of the target shapes and adhered to the drawing instructions (examples in Figure~\ref{fig:expl_results}). Here we present our core findings:

\subsubsection{Drawing Curved Ribbons Was Challenging}

Analysis of the survey responses with a within-population paired t-test \cite{kalbfleisch2012} and qualitative comments indicated that the level of difficulty when surfacing a shape is correlated with the shape's complexity as captured by the curvature variation across it (square/triangle vs cone: mean diff -2.062, std dev: 0.929, t 
= -8.883, p < 0.001; cone vs torus/hemisphere: mean diff -0.625; std.dev 1.258; t = -1.987; p = 0.0033). Participants ranked the planar shapes as easiest to draw, and performed the least amount of corrections when drawing those; and rank the doubly-curved shapes as the most difficult 
and performed the most error corrections when surfacing them (Table~\ref{tab:table1} and supplementary material). From a drawing perspective, the core difference between these shapes is that while the square and the triangle can be drawn using planar ribbons, the hemisphere 
and the torus require using ribbons with non-zero curvature. Notably the cone can be drawn using either curved or planar ribbons (the latter choice requires the ribbons to follow the $k_{min}=0$ direction). Our interviews confirmed that drawing difficulty is indeed linked 
with the need to draw high curvature ribbons when depicting these shapes. Participants' comments supports our interpretation: ``I need to rotate them [the ribbons] and align the ribbons to actually make it curved (P1)'';
 ``When you draw the hole [of the torus], it's kind of, like, a little hard to determine how does the surface should go (P1)'', ``The other problem with the cone is that the surface is curved and the brush is not, the easiest way is to follow the straight lines [the min principal curvature] as oppose to the curve lines (P3).'' 

We conclude that the difficulty stems from the need to draw curved ribbons which exhibit high normal variation. 

\begin{figure}
\centering
  \includegraphics[width=.9\linewidth]{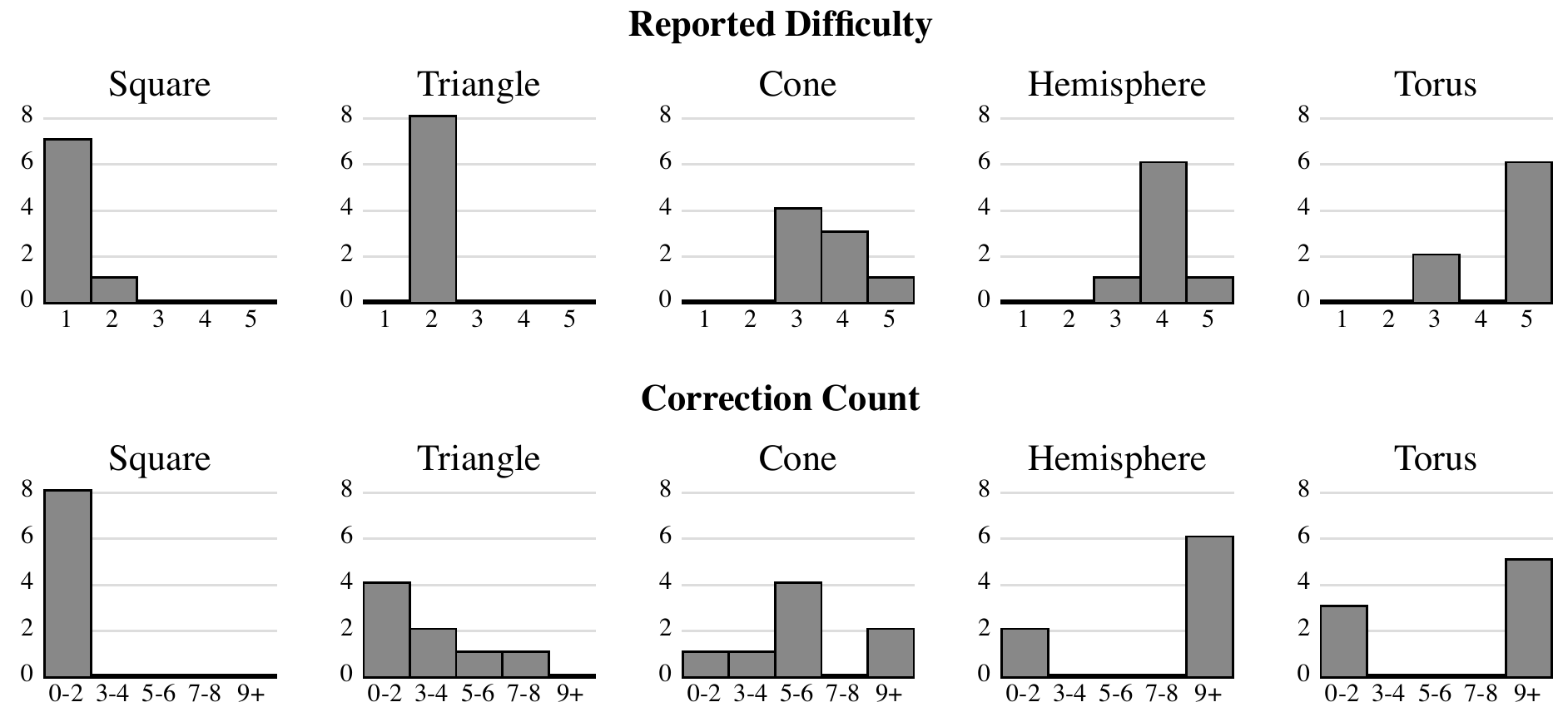}
  \caption{Reported difficulty (top) and correction counts (bottom) per shape.}
  \label{fig:histograms}
\end{figure}

\begin{table}
\centering
\begin{tabular}{|l |c | c|}
\hline
Shape & Difficulty & Corrections  \\
         & avg/median & avg/median\\
         \hline
Square &	1.125	/1 & 	0.750 / 0.5 \\
Triangle &	2.000	/ 2 &	2.625 / 2.0 \\
Cone	& 3.625	/ 3.5 & 	11.125 / 5.5 \\
Hemisphere	& 4.000 /	4	& 12.625 / 13.5\\
Torus &	4.500 /	5	& 21.0 / 9.0\\
\hline
\end{tabular}
\caption{Summary of quantitative findings (formative evaluation): Left to right: shape, average and median difficulty score, average and median error correction count. Shapes ordered top to bottom based on average difficulty.} 
\label{tab:table1}
\end{table}

\subsubsection{The Wrist-Twisting Problem}

Analyzing the annotated video footage as well as qualitative comments gave us a deeper insight as to the roots of the observed curved ribbon drawing difficulty. 
Reviewing the video, we identified critical incidents where users encountered visible difficulty when trying to form long strokes with high normal variation. Specifically, when changing the surface normal by $90^{\circ}$ or more from an initial orientation, they bent and rotated their 
wrists in unnatural directions (see Figure~\ref{fig:painful}, (a) for an illustration of a typical scenario).

These observations are supported by user interview feedback. Users noted:  (when describing drawing the hemisphere) ``The problem is that, when I am drawing, if I would do this [aligning the hand naturally] you see my ribbon is in a completely wrong direction. If my ribbon was pointed flat, [aligning the ribbon horizontally] then the sphere would be a lot easier to trace (P3).''; ``The problem with the cone is that the surface is curved and the brush is not, the easiest way is to follow the straight lines [the min principal curvature] as oppose to the curve lines P3.'';  (describing the torus) ``It has an inner surface, drawing the inner surface is very hard, it is a surface that is constantly curved, so it is difficult to follow the curvature twisting your wrist (P7).''.

\begin{figure}
  \includegraphics[width=\linewidth]{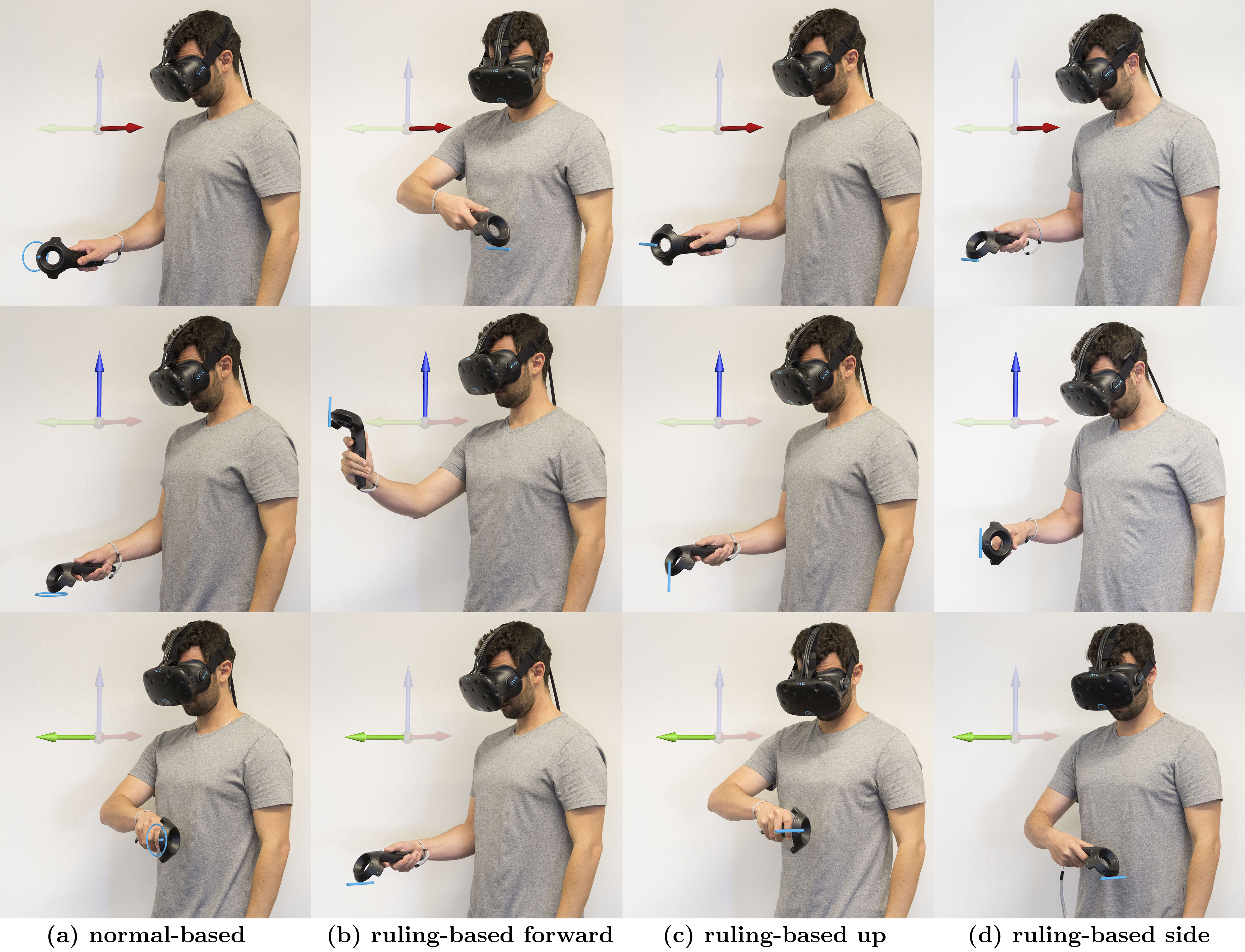}
\caption{(a) Mapping the three major drawing planes (front facing, side facing and horizontal) to corresponding hand-held controller orientations using normal brush. (b-d) Wrist twist minimizing hand and controller orientations necessary to align a local controller axis (b - forward, c - up, d - side) with each of the major global axes (up, side, front).}
\label{fig:controller_pos}
\end{figure}

We further hypothesize that this wrist twisting behavior occurs due to the constraining nature of the current ribbon orientation control, where users are effectively required to align the back of their hands with the surface tangent plane. For a fixed tangent plane, e.g. 
for each of the major planes defined with respect to the user's orientation (Figure \ref{fig:controller_pos}, (a)), one can find a comfortable (straight wrist, low elbow) drawing pose. Drawing curved ribbons however requires continuously changing the hand orientation to correctly depict the intended ribbon normal. This task, as illustrated in Figure~\ref{fig:painful}(a), is likely to require users to gradually bend or twist their wrists into unnatural poses.

We note that there is a unique mapping between the controller's orientation and the user's hand orientation due to fixed button positions on the haptic device: the user cannot simply rotate the current controllers for more comfort, and can only rotate the controller by rotating their hand. 
Wrist flexion and extension, in which the wrist is bent up and down from a neutral pose, is a reasonably comfortable action. 
However, it can only adjust the hand orientation by one Euler angle, 
specifically pitch. The other angle we are interested in, yaw, can only be adjusted by twisting the wrist or bending it sideways. Both 
motions permit a smaller range of angular movement, and are controlled by handle muscles which are not usually as well-developed.
Thus the wrist-twisting behavior identified in our review of video footage cannot be alleviated by users simply choosing a different orientation strategy.  

Based on this finding, we developed a new brush-based interface that attempted to minimize wrist exertion, described in the following section.

\section{StripBrush Design}
\label{sec:design}

Our goal is to find a new control scheme for ribbon drawing which minimizes the amount of unnatural arm and wrist twisting that users need to perform when drawing curved ribbons, without making drawing zero-curvature (flat) ribbons much harder. Our first insight, derived from the exploratory study video footage, is that users can effectively draw ruled ribbons without directly specifying a desired ribbon normal. We further note that the ribbons formed using the normal brush are constructed so that the ribbon rulings are orthogonal to their spatial trajectory. While this property may sometimes be advantageous - for instance, it ensures that, for a fixed ruling length, the ribbon width remains constant - we hypothesize that it can be sacrificed to improve usability without reducing drawing accuracy. Consequently, our key idea is to provide the user with an interface based on an actual ruling instead of a tangential disc, and to allow the angle between these rulings and the controller path to change freely.

We therefore require a way for users to directly specify the ruling directions at each point along the path. We use one of the primary axes of the controller coordinate system (Figure~\ref{fig:controller}) as a ruling direction, since the mental mapping between those axes and the user's controller orientation, and consequently the user's hand gestures, is self-evident. The next important question is which of the three possible axes to use. 
Our experiments, one of which is shown in Figure~\ref{fig:painful}, bottom, indicate that using any of these axes as a ruling, and forming ribbons by sweeping this ruling by moving the controller along a desired path, drastically reduces the amount of wrist twisting compared to the normal-based drawing scheme.
Our choice of axis was therefore made based on two additional design goals. First, we wish to make sure that the drawing process is as easy as possible; second, we wish for an interface that users can easily understand. Figure \ref {fig:controller_pos}, (b-d) illustrates wrist-effort minimizing grip poses for drawing rulings aligned with the major world directions using these three options. The ''side'' and ''up'' orientations require minimal wrist twisting, and allow the users to hold the controller in a {\em power} grip (as defined by ~\cite{Napier}) for all major ruling orientations (Figure \ref{fig:controller_pos}, (c,d)). The power grip distributes the controller's weight along all fingers, making the effort of holding a heavy object easier. In contrast, the ''forward'' axis requires users to switch to a different grip to form vertical rulings, one in which the controller is held vertically; this grip concentrates the object weight on just a couple of fingers, making prolonged drawing more tiring. 
Following the second argument, we note that the drawing process we use is conceptually similar to wall or other surface painting (a connection noted by ~\cite{Rosales:2019} who used the ''fence painting'' metaphor to describe the process that users employ when drawing shapes). Standard paint brushes and rollers are designed for, and held in, the same way as our controller would be when the ''side'' axis is used as a ruling (Figures~\ref{fig:brush},  (a) and  \ref{fig:controller_pos}, (d)). 
We therefore expect the postive transfer would contribute to the learnability of a sideways brush better than the alternatives. 
  
Based on these considerations we converged to the choice of the ''side'' axis, parallel to the controller's top and orthogonal to its handle. Extending the analogy with a paint brush or roller, we center the ruling at the controller's tip, making it symmetric with respect to the ``forward'' axis. Moving the controller using our scheme generates new ribbons that look exactly like paint strips created by moving a roller along the controller path. Controlling the orientation of the ribbons no longer requires wrist twisting motions; instead, the orientation is determined by a rolling gesture, driven by the shoulder and forearm. This alleviates pressure on the wrist and results in natural movements.
We note that the new pre-release of GravitySketch uses  the ``forward'' axis as their mapping choice. Given an obvious lack of documentation we cannot guess the reasoning behind it; we discarded this option for the reasons outlined above.

\subsection{Technical Implementation}
We implemented StripBrush as a Unity application, using the OpenVR SDK and the SteamVR Unity plugin \cite{Unity}. The user interacts with StripBrush using two controllers, one in the dominant hand and one in the non-dominant hand. The dominant hand performs drawing actions; the non-dominant hand controls additional user interface options, such as undo and redo functionality, mimicking the TiltBrush interface. We also provide users with "draw" and an "erase" modes. 

To compute new ribbon points, we sample the dominant hand controller position when the "draw" trigger is engaged, and after the controller has moved an $\epsilon$-distance away from the last ribbon endpoint; we set  $\epsilon$ to empirically match  observed behaviour in other packages. 
We place ruling endpoints at a distances of half a ribbon width away from the controller tip along its ``side'' axis. We then connect consecutive endpoints as discussed in Section~\ref{sec:explorativeStudy} to obtain a ruled ribbon mesh.

\section{Evaluation Methods}
\label{sec:evaluation}
We performed a comparative evaluation of StripBrush against a normal brush implementation (the baseline) using a 2 (Tools) by 8 (Shapes) within-subject factorial design. We asked participants to surface eight different shapes using one drawing tool first, and then the same eight shapes using the other drawing tool. We split our participants into two groups: the first group used StripBrush first, then the baseline; the second group used the baseline brush first, then StripBrush. The order in which participants drew the shapes was randomized.

Two of the shapes were planar (square, circle); two contained singly curved surfaces (cone, cylinder); four were doubly curved (ellipsoid, hyperbolic paraboloid (commonly known as a 'saddle'), sphere and torus.) Three (square, circle, saddle) had open boundaries, the rest were closed; two (cylinder, cone) had sharp features; the rest are smooth. These shapes cover the full spectrum of curvature types.

Our goals were to compare workload (measured using the NASA TLX questionnaire), user perceived accuracy, usability (System Usability Scale, SUS), and user preference for both tools.

\subsection{Apparatuses and Study Setup}
We implemented StripBrush as described in the previous section. To minimize the impact of external factor we re-implemented the normal brush using an identical interface. The only difference between the two interfaces was the brush tip interface (Figure~\ref{fig:brush}): while StripBrush
employed the interface in Section~\ref{sec:design}, the normal brush used the disk, or circle based interface described in Section~\ref{sec:explorativeStudy}.  
During the study, to avoid divulging implementation details, we referred to the two interfaces using their respective signifiers, as {\em LINE} and {\em CIRCLE} tools respectively.
Our study was performed using the HTC Vive VR headset, working inside a drawing area of approximately 3 meters by 3 meters. Participants remained seated during the VR drawing experiments, in a chair placed in the middle of the drawing area.

\subsection{Participants}
We conducted a pilot study with one participant, and conducted the final study reported on below with 17 other participants with different academic backgrounds. 
Six participants were female, and eleven were male. All participants were between 23 and 39 years old. Seven participants had a computer science background, four had an engineering background, and three had a background in visual arts; the rest had backgrounds in dentistry, dance, and business. Four participants had no previous experience using VR. The rest of the participants had less than six months of experience; four of those participants had some previous experience drawing in VR using TiltBrush  \cite{TiltBrush:2019}. Eight participants used StripBrush first, and nine used the normal brush first.

\subsection{Procedures}
Participants were introduced to the features of our VR brush implementation, including a quick introduction on how the LINE (StripBrush) and the CIRCLE (baseline) brush tools worked. We then let participants use the system to get used to the navigation tools and basic drawing, and performed an initial assessment of their depth perception by asking them to perform a simple tasks of drawing three vertical ribbons at the same depth. We then asked participants to perform three more practice tasks, using the undo, delete, and redo features. Again, all participants were able to perform this exercises correctly.

\begin{figure}
\centering
\includegraphics[width=.9\linewidth]{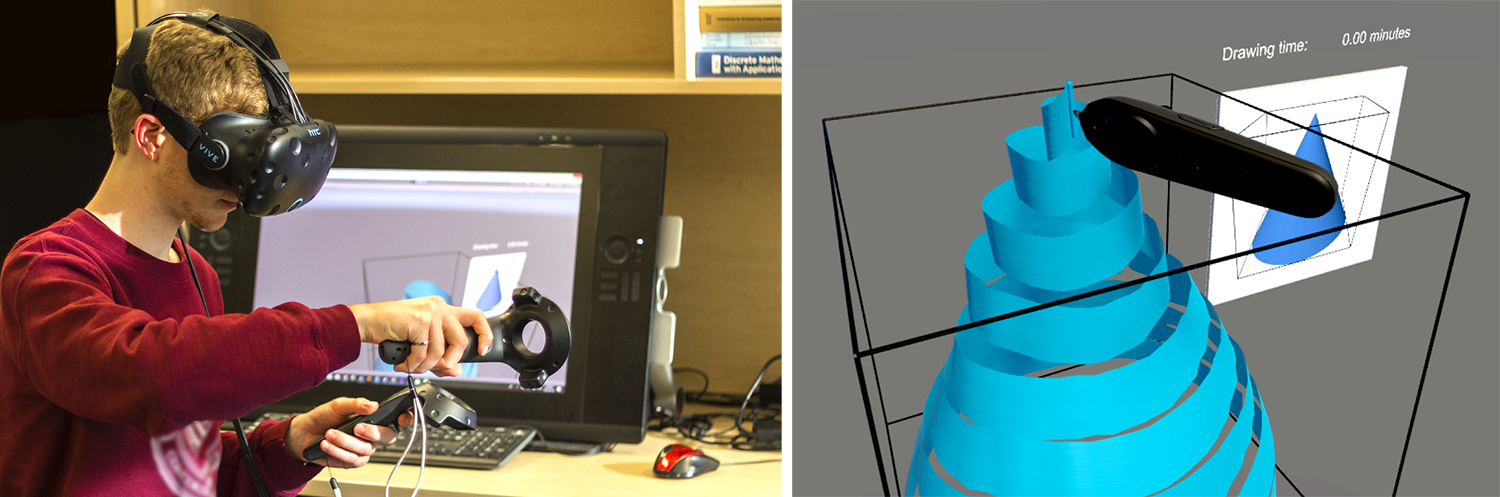}
\caption{Comparative study setup.}
\label{fig:esetup2}
\end{figure}

Once participants felt comfortable using the system, we asked them to perform the main study task inside the VR environment. We showed participants a 2D reference image of a surface inscribed in a bounding box, and presented them in the virtual reality space with a three-dimensional bounding box with the same proportions of the one on the image (Figure~\ref{fig:esetup2}, right). We then asked the participants to depict the reference surface as accurately, or as cleanly, as possible. Participants were shown a timer bar set to 3 minutes; we explained that this was a typical time for drawing these surfaces, but that they should take more or less time as needed. We again emphasized that the goal was to produce accurate descriptions of the reference shapes.

After drawing each shape, we asked participants to verbally answer their level of agreement (using a 5 point Likert scale) with the following sentence: ``drawing this shape was easy with the [LINE/CIRCLE] tool''. We then confirmed their response by reading their agreement level from the Likert scale. The recording was done orally since the participants were wearing the VR headsets at that point. 

We repeated this task for each of the eight different shapes using the first tool; after that, we asked participants to take off the VR headset and answer (on a laptop) the weighted NASA TLX workload survey~\cite{hart1988development}, which assessed their perceived workload when using the tool, and a System Usability Scale (SUS) survey ~\cite{albert2013measuring} which measured users assessment of the tool's usability. We then repeated the eight drawing tasks and collected similar survey responses for the second tool. 
Participants answered a pre-task demographic survey and a post-task survey on their experience during the tasks. 
After completing both sets of tasks, we conducted follow-up interviews with the participants  focusing on usability, and asked them to elaborate on their answers on the usability survey and compare the two tools they used from a usability perspective. 
During the task sessions, our software counted the number of correction (undo/redo/delete) operations and runtime, and saved the completed drawings.

\begin{figure}
\centering
  \includegraphics[width=\linewidth]{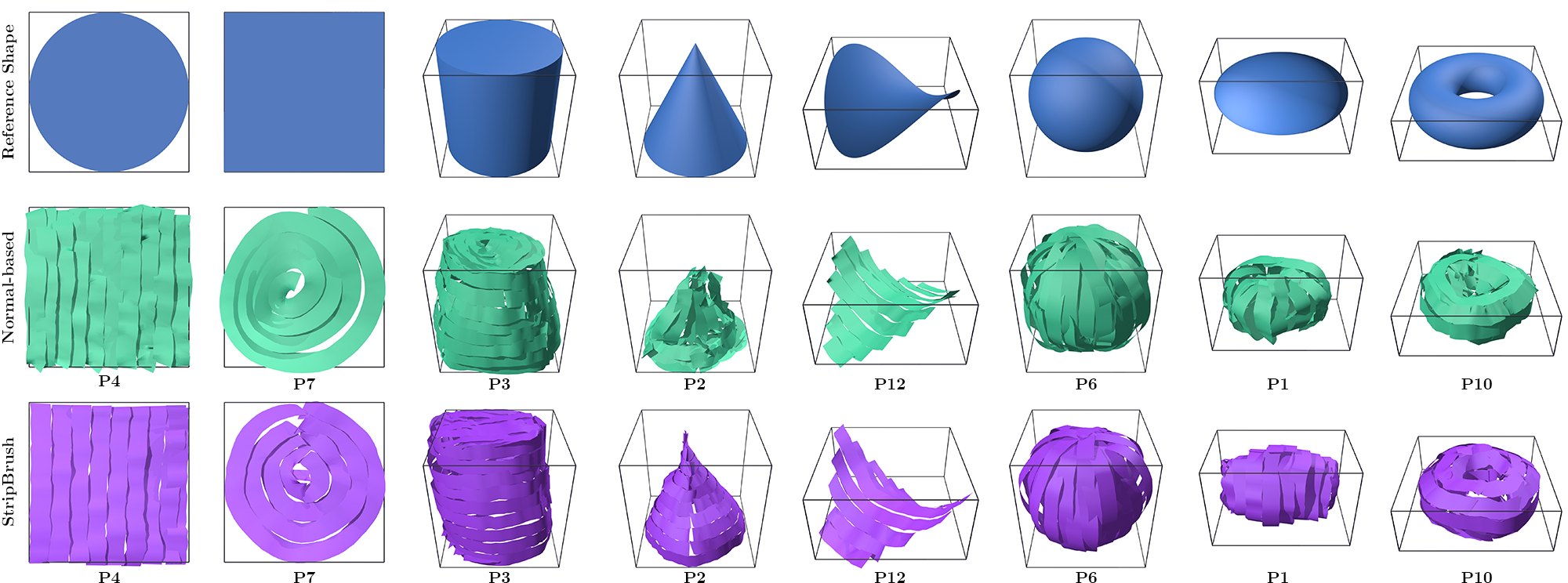}
\caption{Reference shapes used in the study and representative pairs of participant drawings created using normal brush (green) and StripBrush (purple).}
\label{fig:study2}
\vspace{-10pt}
\end{figure}

\section{Results}
\label{sec:results}

We summarize the study results below. Example user drawings created using both tools are shown in Figure~\ref{fig:study2}. All results and study data are provided as supplementary material. In the data analysis below, we recall that all quantitative evaluations we performed 
are within-group; in other words, the information we need to assess is the difference between the quantitative metrics provided for the two tools by the same user (SUS,TLX), and for questions and propertied measured per drawn shape, the difference between the corresponding 
quantities for the same user and for the same reference shape. We visualize these differences in the figures below, and use a paired t-test \cite{kalbfleisch2012} when comparing samples from the normal brush and StripBrush populations when appropriate. As is common in these cases, we report the mean of differences and the standard error of mean differences only.

Analysis of our results reveals strong statistical evidence that StripBrush is less physically demanding than the normal brush, allows participants to describe shapes more accurately than the normal brush, and is rated by participants as significantly more usable overall as 
well. Users found StripBrush to be less frustrating than normal brush, and also made fewer corrections when using it. At the same time, we observe that there is no statistically significant change between StripBrush's and the normal brush's performance on a range of other 
comparative measures. From this, we conclude that StripBrush offers a clear improvement on normal brush interfaces in terms of physical exertion, accuracy, frustration and usability, with no significant downsides.

\begin{figure}
\centering
  \includegraphics[width=\linewidth]{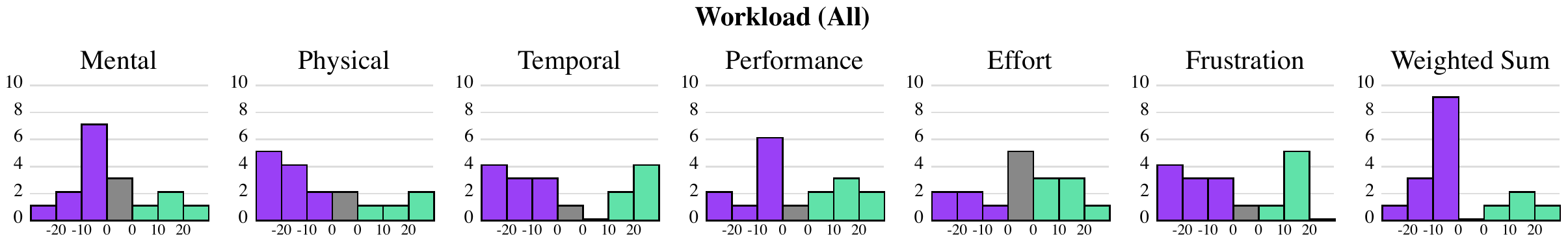}\\
  \includegraphics[width=\linewidth]{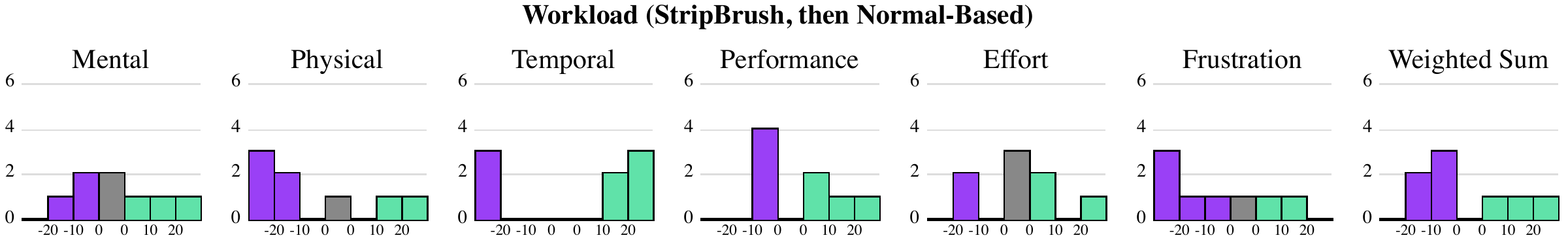}\\
\includegraphics[width=\linewidth]{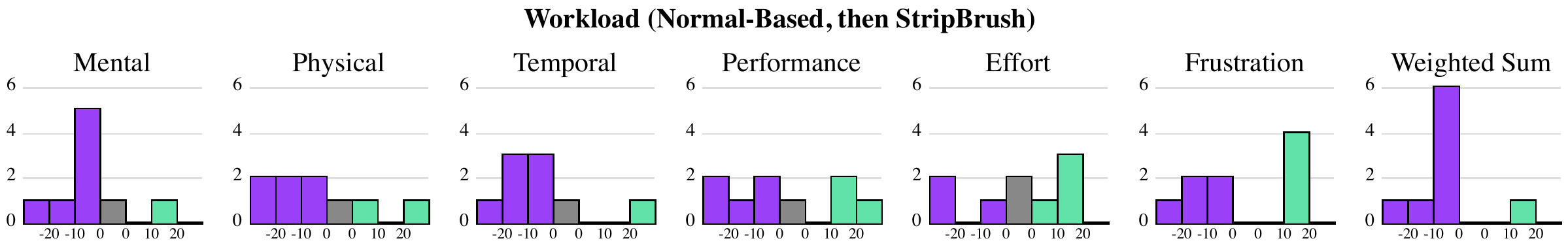}\\
\includegraphics[width=\linewidth]{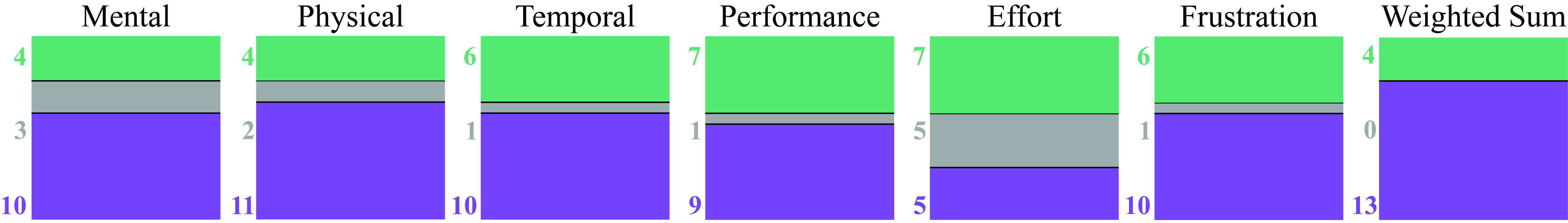}
\caption{Summary of workload TLX feedback. All values shown are differences between corresponding values for StripBrush and normal brush interfaces. Lower (negative) numbers indicate better performance by StripBrush, higher ones indicate better performance by the alternative. (top) All participants; (row 2) participants who used the StripBrush tool first; (row 3) those who used the normal brush first. Shading (purple vs green) highlights the halfspaces which correspond to better performance of StripBrush and the normal brush respectively. (bottom row) Participant preference summary: (purple) prefer StripBrush, (green) prefer normal brush, (grey) neutral.}
\label{fig:workload}
\end{figure}

\subsection{StripBrush is Less Physically Demanding Than Baseline}
We summarize our workload (NASA-TLX) study findings in Figure~\ref{fig:workload}. 
Using the NASA-TLX scale, participants deemed StripBrush to be less physically demanding than the traditional normal brush (t=-1.921; p=0.036, diff=-13.529, std=29.035; 95\% confidence interval: (-28.458, 1.399)). 
We therefore find strong evidence that our system is less physically demanding than the normal brush approach. We analyzed the impact of tool access order on the perceived workload by separating the findings for the group that used StripBrush first (Figure~\ref{fig:workload}, row 2) and for those who used it second (Figure~\ref{fig:workload}, row 3). In general, one may expect workload to reduce as participants gain experience. 
We note that the preference for StripBrush over normal brush among participants who used StripBrush as their second tool (row 3) is stronger across all workload indicators; however, even among the first group we observe a clear difference in physical workload perception, with five participants scoring StripBrush as being less physically demanding, two scoring the normal brush as such, and one scoring them as equal (for the second group the split is six to two with one neutral score). 

\begin{figure}[t]
\centering
  \includegraphics[width=\linewidth]{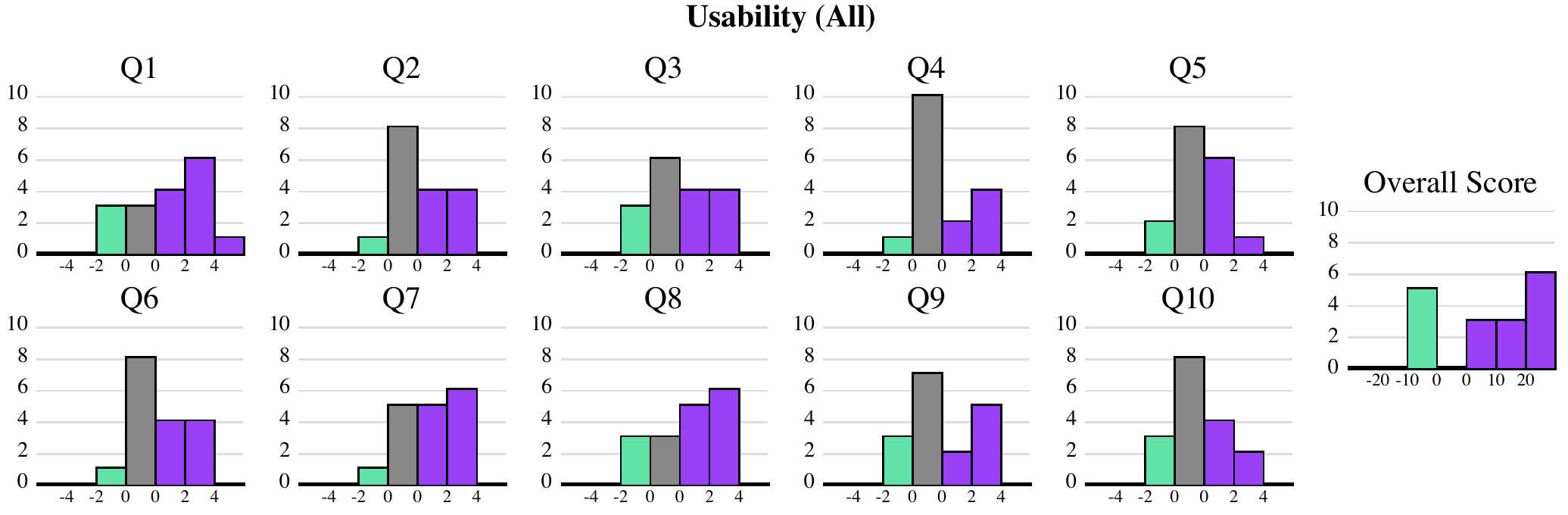}\\
  \includegraphics[width=\linewidth]{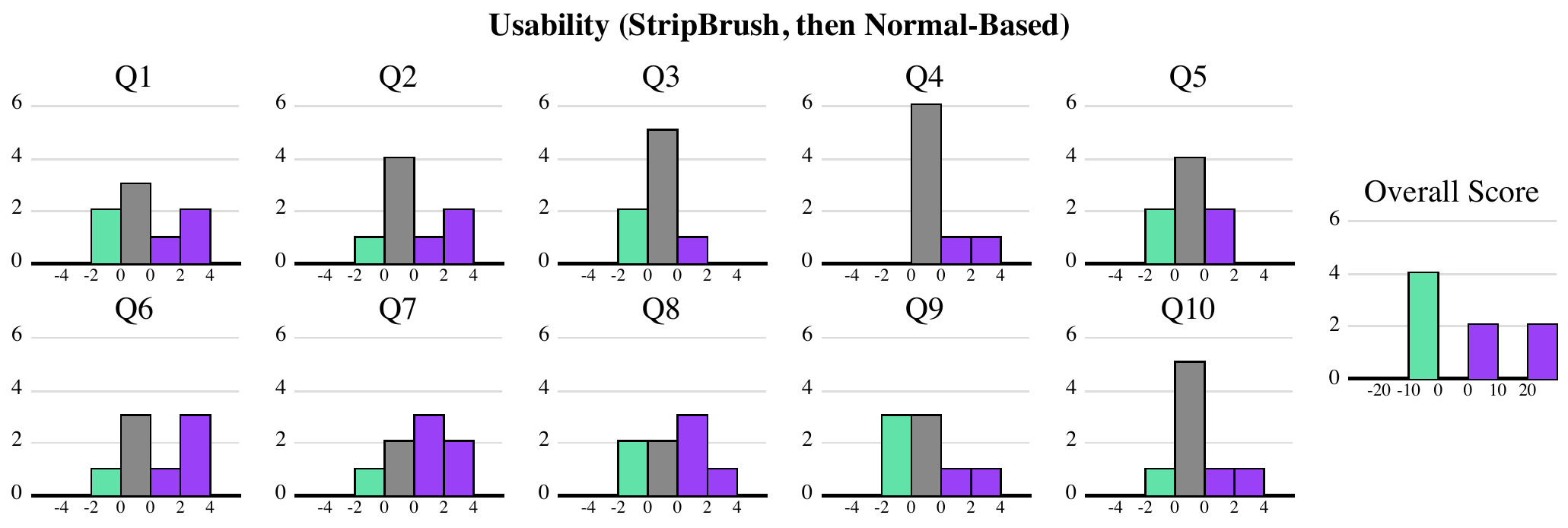}\\
\includegraphics[width=\linewidth]{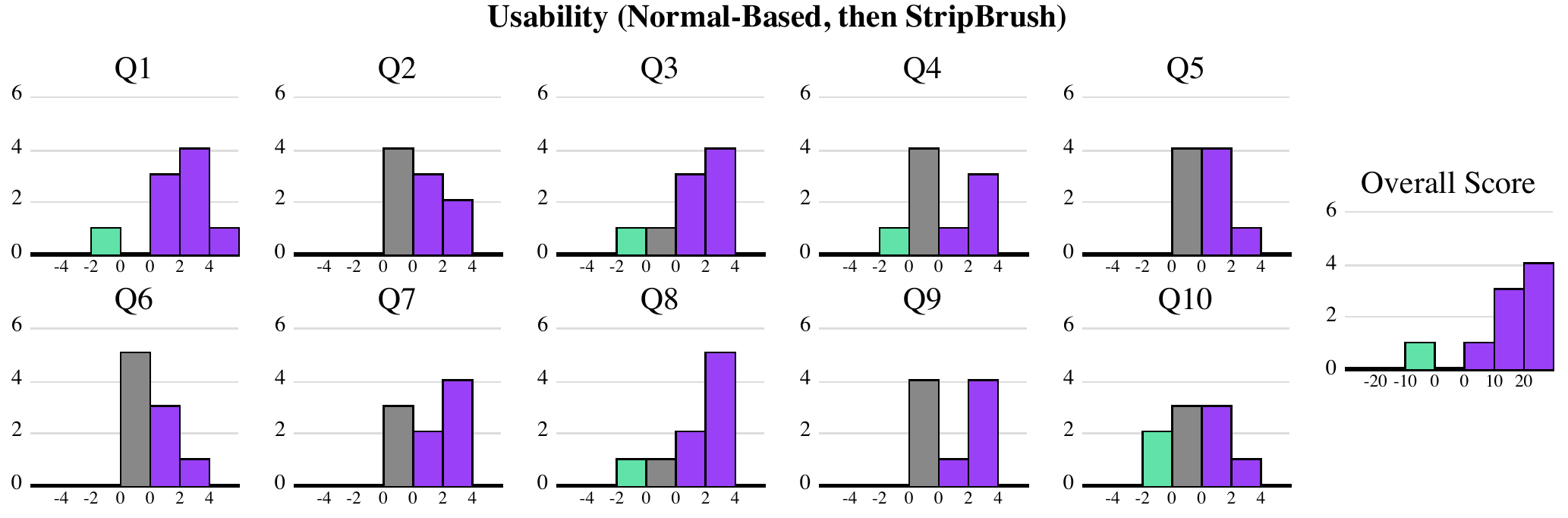}\\
\includegraphics[width=\linewidth]{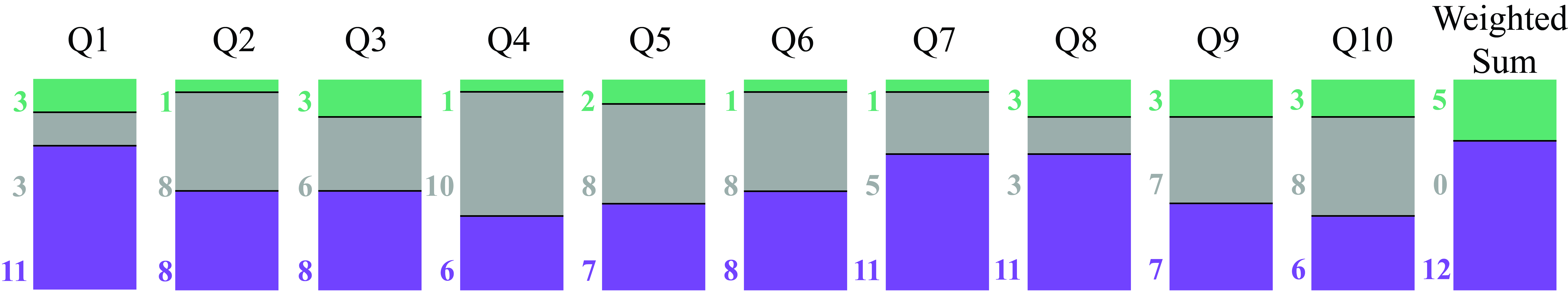}
\caption{Summary of usability (SUS) feedback. All  values shown are differences between corresponding values for StripBrush and normal brushes. Higher (positive) numbers indicate better performance by StripBrush: (rows 1-2) all participants;  (rows 3-4) participants who used the StripBrush tool first; (rows 5-6) those who used the normal-based tool first. Last row shows aggregate preferences. }
\label{fig:usability}
\end{figure}

We compare the overall participant perceived usability of StripBrush and normal brush interfaces using SUS (Figure~\ref{fig:usability}). Based on our collected data, specifically the combined usability score (last column), participants find StripBrush more usable than the normal brush (t=3.427; p=0.002, diff=13.882 std=16.484, confidence interval: (5.407, 22.358)). We conclude that there is strong statistical evidence that our interface is significantly more usable than normal brush interfaces. As with workload, the preference is stronger in the group that used StripBrush second (Figure~\ref{fig:usability}, rows 5-6) with 8 participants scoring StripBrush as more usable overall (last column), compared to one scoring the normal brush as more usable); in the group that used StripBrush first (rows 3-4) the overall preferences (last column) were evenly split. 

\begin{figure}[t]
\centering
\includegraphics[width=\linewidth]{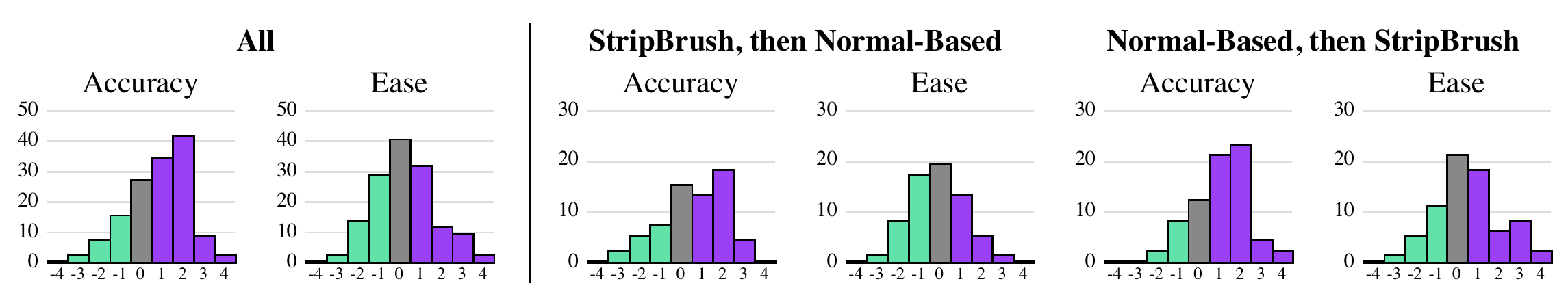}
\caption{Feedback collected during user drawing sessions on a per-participant, per shape basis. The values shown are differences between corresponding (same participant, same shape) scores for StripBrush and normal brushes. Higher (positive) numbers indicate better performance by StripBrush, compared to the normal brush. (left) Difference in perceived usability and accuracy across {\bf all} participants. (center, right) Breakdown of responses separating participants who used the StripBrush tool first (center) and those who used the normal brush tool first (right).}
\label{fig:individual_evaluation}
\end{figure}

\subsection{StripBrush Enhances Perceived Accuracy of Drawing}

\begin{figure}
  \begin{center}
      \includegraphics[width=.9\linewidth]{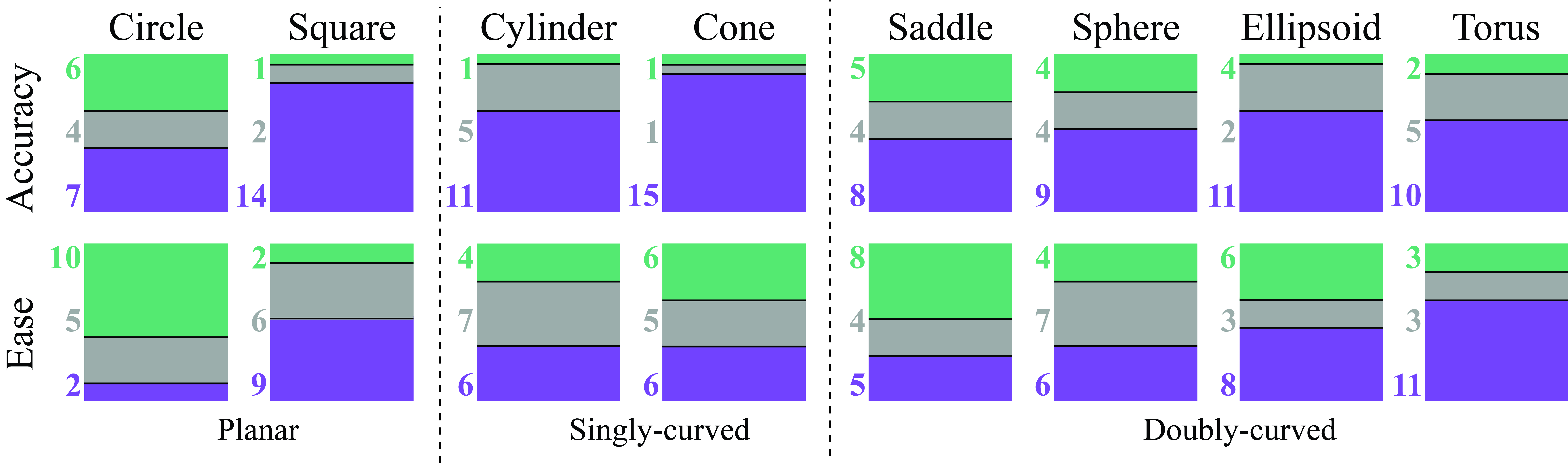}
        \end{center}
\caption{User perceived easiness and accuracy feedback scores per reference shape, expressed as preference values. Green: participants preferred normal brush as being easier/more accurate than StripBrush. Grey: no preference. Purple: StripBrush was preferred for ease/accuracy. We see that participants clearly prefer StripBrush for accuracy, and find it at least as equally easy to use as normal brush.}
\label{fig:chart_easy_by_shape}
\end{figure}

After completing all drawings, we showed participants each reference shape and the drawings they created for this shape using the two tools. We asked them to rate how strongly they agreed with the statement "This drawing accurately represents the surface on the image", using the Likert scale, for each of their drawings. 
The differences between the scores are summarized in Figure~\ref{fig:individual_evaluation}. We find that, on average, participants find results created using StripBrush significantly more accurate (t=6.863; p<0.001; diff=0.831, std=1.412; conf. interval: (0.591, 1.070)) than those created using the normal brush, and conclude that there is strong statistical evidence that users feel they are able to more accurately depict their intended shapes using StripBrush than using the normal brush. The breakdown based on tool use order (Figure~\ref{fig:individual_evaluation}, right) shows that both groups perceive their StripBrush drawings as more accurate, with the preference being much stronger for the group who used it second.

\setlength\columnsep{3pt}
\begin{wrapfigure}[4]{l}[0pt]{.17\linewidth}
 \vspace{-12pt}
  \begin{center}
   \includegraphics[width=\linewidth]{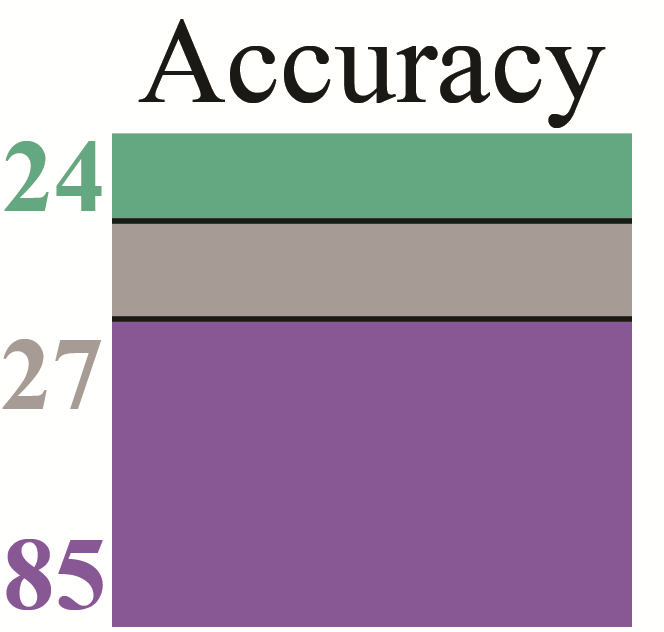}
  \end{center}
\end{wrapfigure}
In addition to score differences, we computed preference data (counting the number of times each participant scored one interface higher than the other for each individual shape drawn). As summarized in the inset 85 responses indicated that StripBrush drawings were more accurate than those drawn with normal brush; 27 responses indicated that StripBrush and normal brush drawings were equally accurate; 24 responses indicated that normal brush drawings were more accurate. A per-shape breakdown of these preferences is shown in Figure~\ref{fig:chart_easy_by_shape}. These numbers further reinforce the observation that using  StripBrush users can more accurately draw their intended shapes than when using the normal brush.

We also asked participants to rate, for each shape, whether they thought it was easy to draw it with Stripbrush, and separately whether they thought it was easy to draw it with normal brush (these responses were collected immediately after they drew each shape). We  phrased the question as "Drawing this shape was easy with the (LINE/CIRCLE) tool." Participants were asked to rate their agreement with this statement on the Likert scale.
The differences between the scores are summarized in Figure~\ref{fig:individual_evaluation}. We find that, on average, participants find StripBrush significantly easier to use on a per-shape basis (t=1.666; p=0.049; diff=0.206, std=1.441; confidence interval: (-0.038, 0.450)) compared to normal brush. 
The breakdown based on tool use order (Figure~\ref{fig:individual_evaluation}, right) shows that while the participants who used StripBrush first see the normal brush as slightly easier to use; those who used it second strongly prefer StripBrush.

\setlength\columnsep{3pt}
\begin{wrapfigure}[4]{l}[0pt]{.17\linewidth}
 \vspace{-12pt}
  \begin{center}
   \includegraphics[width=\linewidth]{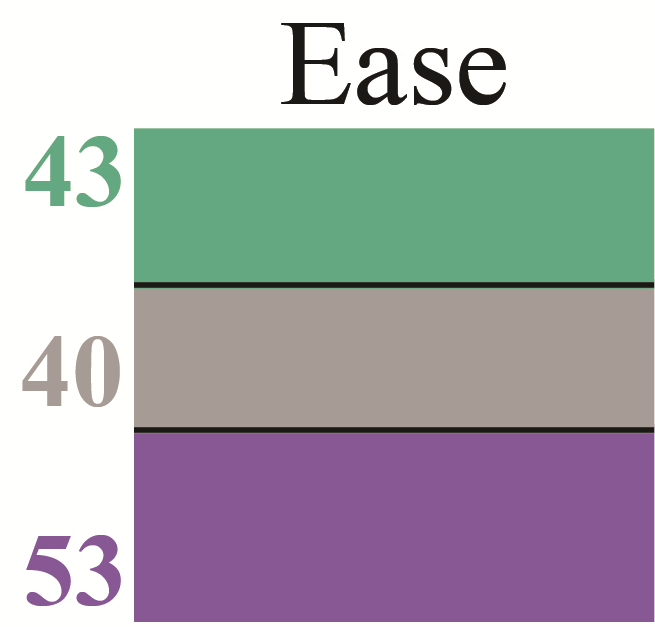}
  \end{center}
\end{wrapfigure}
In addition to score differences, we computed preference scores. Overall, 53 responses indicated that StripBrush was easier to use than normal brush; 40 responses indicated that StripBrush and normal brush were equally easy to use; 43 responses indicated that normal brush was easier to use than Stripbrush (see inset). A per-shape breakdown of these preferences is show in Figure~\ref{fig:chart_easy_by_shape}.

Participant interviews reinforce our conclusions. Oral feedback included comments such as: ``Line is simple, there is too much effort for the circle, I needed to keep my hand steady, and the circle makes more noise'' (P5), ``I didn't manage to draw with the circle until [having] more practice. The line is very easy to learn; with the circle it's very easy to get the ribbon wrongly twisted because I need to rotate my wrist'' (P9), ``The line tool is much more controllable'' (P17), and ``The line tool felt more technically precise when you start and stop, you can do very clean paths'' (P3). Several users commented on cases where the normal-based tool may be more suitable:  ``To make a planar curve, the circle is easier, that will be my only preference for that tool'' (P10), and ``Circle has two degrees of freedom, line tool is much easier, most of the images are easier with the line tool except for the planar circle'' (P11)

These observations align with our conclusion that StripBrush is generally less taxing and is most beneficial for drawing curved shapes, the most common use scenario for such tools~\cite{TiltBrushRepo:2019}. They also highlight the improved drawing accuracy it provides.

\subsection{Performance}

\begin{figure}[h!]
\centering
\includegraphics[width=.85\linewidth]{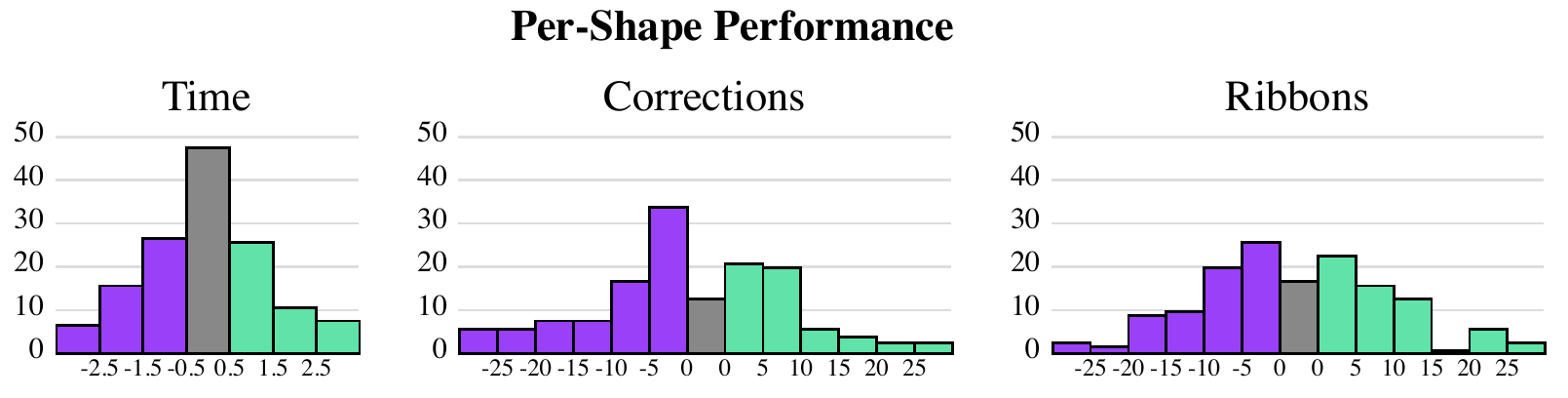}
\caption{Performance measurements collected during study drawing sessions on a per shape basis shown as differences between corresponding values for StripBrush and normal brush.  Lower (negative) numbers indicate better performance by StripBrush.  Left to right:  runtime (seconds), correction count, ribbon count. There difference in values when breaking the responses based on tool assessment order is negligible. Both methods perform equally well for these metrics.}
\label{fig:performance}
\end{figure}

When directly comparing user performance using StripBrush vs. the alternative, on a per-shape, per-user basis, participants take the same amount of time on average to describe their target shapes and use on average the same number of strokes.   We conclude that StripBrush and the normal-based approach perform equally well in terms of time and ribbon count. Participants perform  fewer correction operations (undo, redo, delete) when using StripBrush (Figure~\ref{fig:performance}). The difference in correction count was statistically significant (diff: -2.235, std=11.368, CI (-4.163, -0.307); t=-2.293, p=0.012). We saw no perceivable difference when breaking the performance down into groups based on tool use order.

\setlength{\abovecaptionskip}{3pt}

\section{Discussion}
\label{sec:discussion}

We hypothesized that the design of StripBrush would reduce physical exertion, and this is clearly shown in the results. Users also found our tool easier to use than the normal-based brush, and we were surprised that users perceived the tool as being more {\em accurate} than 
the normal brush. We hypothesize that the ease-of-use of StripBrush, in addition to the reduction of physical exertion, is because it presents a familiar metaphor: users are familiar with a paint roller and know how it works, and can intuitively apply this metaphor to VR 
spaces. We hypothesize that the higher accuracy score is due to two factors. First, participants using a tool that they find easier are more likely to relax when using it, and hence more likely to produce a result that they perceive as accurate. Second, StripBrush grants 
users a finer degree of control over generated ribbons. Consequently, they are able to more precisely achieve the idea that they ideate in their head, and hence perceive the tool as more accurate.

As the target sufaces used in our evaluation cover a wide variety of curvature characteristics, the results from our controlled evaluation (Section \ref{sec:evaluation}) can serve as useful empirical grounds to estimate advantages and limitations of StripBrush in real 
drawing settings when the user draws more general shapes. As the catalogue of a popular spatial sketch sharing platform \cite{TiltBrushRepo:2019} indicates, artists and VR drawing hobbyists tend to draw more organic subjects that require the use of curved ribbons to 
successfully surface. We expect StripBrush to have context-specific advantages on these designs, rendering it more promising than what can be reported in a controlled study.

At the same time, while StripBrush performs best when drawing curved ribbons, the normal brush - or another brush entirely - may be easier to use for planar surfaces. It would therefore be useful to consider hybrid solutions where the system can adaptively transition 
between the two interaction modes by detecting and further predicting the curvature of the target surface intended by the user, given the drawing context. Further taking the constraint relaxation as a design space, one may explore the impact of relaxing the constraint 
further than what is designed for StripBrush where a pen tip can be simply represented as a \emph{point} at the extreme end of the circle-line-dot continuum, and the ribbon orientation is derived from contextual cues.

\section{Conclusions and Future Work}
\label{sec:conclusions}

Our paper studied the usability challenges in ribbon-based VR brush drawings. Our study concluded that users experience both difficulty and physical discomfort when drawing ribbons with large surface normal variation. We proceeded to propose an alternative brush tip 
interface, StripBrush, which as validated by our comparative study, reduces the physical effort required to draw 3D shapes, improves drawing accuracy and overall usability, and maintains a range of other performance indicators on par with state-of-the-art normal 
brush interfaces.
  
Our work has a number of exciting followup avenues. Designing and implementing a hybrid system like the one imagined in Section~\ref{sec:discussion} is a major and promising undertaking. As a stepping stone toward this goal it would be interesting to come up with concise user guidelines for when they want to switch between tools. Finally, while our tool reduces the physical effort involved in drawing, it does not eliminate it. Further research could explore other means to further reduce this 
effort.

\section*{ACKNOWLEDGMENTS}
We are deeply grateful to Jonathan Griffin and Nico Schertler for their help on this project. This work has been supported by NSERC and CONACYT.

\appendix
For completeness, we provide a short informal primer on the geometric and differential properties involved in ribbon surfaces and the shapes users create with them. Readers interested in a more rigorous presentation of these properties can refer to~\cite{doCarmo}.
Given a differentiable surface $S$ in $R^2$ defined as a parametric function of two parameters $S(u,v)$, the tangent plane of the surface is spanned by the partial derivative vectors $\delta S_u = \frac{\delta S(u,v)}{\delta u}$ and $ \delta S_v = \frac{\delta S(uv)}{\delta 
v}$. The normal to the surface $n$ is then defined as $n = \frac{\delta S_u}{\| \delta S_u\|} \times \frac{\delta S_v}{\| \delta S_v\|}$. 
Informally, the {\em curvature} of a curve $C$ at a point $p$ is defined as the reciprocal of the radius of the circle that most closely conforms to the curve at the given point (where a straight line is viewed as a circle of infinite radius). The {\em normal curvature} of 
a surface at a point $p$ with respect to a tangential direction $t$ is defined as the curvature at $p$ of a curve formed by intersecting the surface with a plane that contains $p$ and whose normal is orthogonal to both $t$ and the surface normal at $p$. The maximal 
$k_{max}$ and minimal $k_{min}$ curvatures at each point $p$ are defined as the maximal and minimal values of normal curvature at this point. The curvature directions that correspond to these values are orthogonal, and are referred to as the principal curvature directions.  
The surface curvature at any given point can be characterized based on the properties of the minimal and maximal curvatures (positive, negative, zero). Notably, the normal curvature of a ruled surface along the ruling direction is always zero.

% BALANCE COLUMNS
\balance{}

% REFERENCES FORMAT
% References must be the same font size as other body text.
\bibliographystyle{SIGCHI-Reference-Format}
\bibliography{bibliography}

\end{document}